\documentclass[
preprint,
 amsmath,amssymb,
 aps,
prb,
]{revtex4-2}

\usepackage{graphicx}
\usepackage{dcolumn}
\usepackage{bm}
\usepackage{hyperref}
\usepackage{upgreek}

\usepackage{caption}[style=base,justification=justified]

\begin{document}


\title{Giant universal conductance fluctuations in \\ 
the antiferromagnetic topological insulator MnBi$_2$Te$_4$
}


\author{Michael Wissmann$^{1,2,3}$}
\author{Joseph Dufouleur$^{1}$}
\author{Louis Veyrat$^{1,6}$}
\author{Anna Isaeva$^{4,5}$}
\author{Laurent Vila$^{3}$}
\author{Bernd Büchner$^{1,2}$}
\author{Romain Giraud$^{1,3}$}
\email{romain.giraud@cea.fr}
\affiliation{$^1$Leibniz Institute for Solid State and Materials Research, IFW Dresden, Helmholtzstrasse 20, 01069 Dresden, Germany}
\affiliation{$^2$Department of Physics, TU Dresden, D-01062 Dresden, Germany}
\affiliation{$^3$Université Grenoble Alpes, CNRS, CEA, Spintec, F-38000 Grenoble, France}
\affiliation{$^4$Department of Physics, Technical University of Dortmund, D-44227 Dortmund, Germany} 
\affiliation{$^5$Research Center "Future Energy Materials and Systems", D-44780 Bochum, Germany}
\affiliation{$^6$Laboratoire National des Champs Magnetiques Intenses LNCMI, CNRS-INSA-UJF-UPS, UPR3228, F-31400 Toulouse, France}
 




\date{\today}

\begin{abstract}

\textbf{Intrinsic magnetic topological insulators can host quantum states with quantized magneto-electric responses, such as the axion and Chern insulators states evidenced in ultra-thin MnBi$_2$Te$_4$ films. 
Yet, whereas quantization is investigated thoroughly, transport properties related to the phase of charge carriers remains unexplored. 
Here, we study quantum coherent transport in mesoscopic Hall bars fabricated from thick exfoliated MnBi$_2$Te$_4$ flakes, and reveal the longest phase-coherence length $L_\varphi$ ever observed in a mesoscopic magnet ($L_\varphi\approx 500$~nm at 1K), associated to 2D topological surface states. In the fully-coherent regime, significant non-local contributions to quantum interference up to the micron scale lead to giant-amplitude universal conductance fluctuations ($\delta G^{rms}\approx 20$~e$^2$/h). In the self-averaging regime, the statistical properties of conductance fluctuations confirm the 2D nature of quantum interference and different dephasing mechanisms are identified, as due to either magnetism or magnetic flux through coherent loops. Remarkably, the weak decoherence in magnetic topological insulator nanostructures show their potential to realize novel quantum spin interferometers based on dephasing by local magnetic textures at liquid-helium temperatures.}

\end{abstract}

\maketitle



The magnetization dynamics in nanostructures is a key limiting factor to quantum interference effects 
evidenced by charge transport measurements, since inelastic scattering in a magnet can severely reduce the decoherence time $\tau_\varphi$ and the related phase-coherence length $L_\varphi$. Locally-fluctuating magnetic moments usually destroy quantum coherence \cite{Benoit1992,Pierre2003}, and so do collective excitations (magnons) in a ferromagnet, so that quantum corrections to the conductance are vanishingly small in ferromagnetic-metal nanostructures, with $L_\varphi<30$~nm at $T=30$~mK, and the temperature dependence of decoherence is stronger than in normal metals \cite{Hong1995,Aprili1997,Kasai2003,Lee2004,Wei2006,Bolotin2006}. In diluted magnetic semiconductors, finite conductance fluctuations arise from the giant spin-split electronic band induced by local impurities \cite{Jaroszynski1995}, however without long-range magnetic order. 
A significant increase of $L_\varphi$ at very low temperature, by about an order of magnitude, was evidenced in nanostructures of single-crystalline ferromagnets with a large anisotropy \cite{Wagner2006,Vila2007,Neumaier2007}, for which magnetic fluctuations are frozen at very low temperatures due to the magnetic band-gap opening. When $L_\varphi$ becomes comparable to the size of magnetic domains or domain walls, quantum conductance fluctuations in magnetic nanostructures were predicted \cite{Tatara1997,Lyanda-Geller1998,Adam2006}. Indeed, in the diluted ferromagnetic semiconductor (Ga,Mn)As, two additional contributions to dephasing related to magnetism were evidenced, as induced by local magnetic textures \cite{Vila2007} or by Berry phase effects within magnetic domains \cite{Granada2015}. 
Yet, mesoscopic magnets are scarce and, to date, decoherence remains an important limitation to realize quantum interference devices based on magnetic materials, even for state-of-the-art mesoscopic devices built from diluted magnetic semiconductors. 

Recently, the antiferromagnetic topological insulator MnBi$_2$Te$_4$ has gained increasing attention as a platform to realize quantized topological phases controlled by magnetic order \cite{Li2019,Zhang2019,Otrokov2019,Deng2020,Liu2020,Lee2021,Qiu2025}, in particular the quantum anomalous Hall state, of interest for quantum metrology \cite{Poirier2025,Rodenbach2025}. 
Similarly to diluted ferromagnetic semiconductors, intrinsic magnetic topological insulators are crystalline magnets with a large uniaxial anisotropy. In addition, their homogeneous magnetization and uniaxial anisotropy give larger energy gaps, for both their magnetic excitation spectrum and their electronic band structure. This situation is favorable to quantum interference phenomena and, indeed, universal conductance fluctuations (UCF) related to magnetism were evidenced in large Hall bars patterned from ultra-thin MnBi$_2$Te$_4$ films grown by molecular-beam epitaxy, yet with a rather small amplitude of UCF for long conductors, in the 10$^{-2}$e$^2/h$ range \cite{Andersen2023}. 
In this work, we evidence the fully-coherent regime of universal conductance fluctuations in micron-size nanostructures patterned from exfoliated high-quality 
MnBi$_2$Te$_4$ single crystals, 
with an exceptionally-large UCF amplitude for a mesoscopic magnet, much larger than the conductance quantum $G_0=$e$^2/$h. For short-length conductors ($L<L_\varphi$), the root-mean-square amplitude of conductance fluctuations $\delta G^{rms}$ is measured as large as 20~G$_0$, and quantum interference is mostly related to non-local contributions of extended paths associated to 2D topological surface states, with a phase-coherence length enhanced at the micron scale. Specific to intrinsic magnetic topological insulators, mesoscopic transport is found very robust in temperature, a direct signature of the weak decoherence of 2D spin-textured Dirac fermions, allowing us to give the first observation of conductance fluctuations in a mesoscopic magnet at 4.2K. 


\newpage
\textbf{The emergence of giant universal conductance fluctuations}

In a mesoscopic conductor, the amplitude of quantum corrections to the classical conductance depends on its relative size with respect to the phase-coherence length $L_\varphi$, the length scale over which the phase memory of the wavefunction is preserved \cite{Lee1987}. Due to microscopic disorder, quantum interference between coherent paths leads to reproducible conductance fluctuations that can be observed under an applied magnetic field. For short conductors ($L\lesssim L_\varphi$), the UCF amplitude in the fully-coherent regime is independent of the length $L$ and $\delta G^{rms}\approx$e$^2/$h. This is the signature of universal conductance fluctuations, independent of microscopic disorder details. For long conductors ($L>L_\varphi$), the amplitude is reduced due to the averaging of incoherent sections of size $L_\varphi$. In order to investigate the influence of the sample planar geometry on self-averaging, the universal conductance fluctuations were investigated in two devices (A,B), with a different cross-section and for three different lengths (for details, see Fig.~\ref{fig1}a,b).
We first evidence the length dependence of the magneto-conductance measured at $T=1$~K, by applying a magnetic field $B_\perp$ perpendicular to the sample plane, as shown in Fig.~\ref{fig1}c,d).  
The evolution of the classical magneto-conductance is typical for the magnetization behavior of an A-type AFM with an out-of-plane uniaxial anisotropy \cite{Tan2020}. It gives a plateau in the antiferromagnetic (AFM) regime, then a jump at the spin-flop transition at $B_\perp\approx3$~T, followed by a linear increase due to the continuous rotation of magnetic domains in the canted-AFM (c-AFM) regime, and finally a slope change occurs at the saturation field $B_\perp^{sat}\approx7.5$~T above which the magnetization is uniform (UM regime) and aligned along the applied field. 
  
In addition, there is a clear contribution of quantum interference to the magneto-conductance, 
with an increased UCF amplitude when the length is reduced, as expected for a mesoscopic conductor in the self-averaging limit ($L_\varphi<L$). The fully-coherent regime, with $\delta G^{rms}\approx$~e$^2$/h, is already achieved at $T=1$~K for micron-long conductors, which is a direct signature of the phase-coherence length enhancement in MnBi$_2$Te$_4$. 
By contrast, sub-micron (Ga,Mn)As nanostructures show vanishingly-small UCF for all lengths at 1~K, with a short coherence length $L_\varphi\approx30$~nm \cite{Wagner2006,Vila2007}. 
Remarkably, giant-amplitude UCF are observed for short-length conductors, in the super-coherent regime ($L_\varphi \gg L$), with an amplitude $\delta G^{rms}\approx 20$~e$^2$/h that largely exceeds the conductance quantum limit. In this regime, non-local contributions to quantum interference lead to a larger amplitude of conductance fluctuations, as known for non-magnetic mesoscopic conductors \cite{Lee1987,Skocpol1987,Benoit1987}.

\textbf{Nature of quantum interference and robust quantum coherence}

After removing the classical magneto-conductance background, details of the quantum magneto-fingerprint curves reveal the different nature of dephasing associated to the three different magnetic states (Fig.~\ref{fig1}e,f). For large magnetic fields (UM regime), reproducible UCF traces and retraces are observed, as due to orbital dephasing related to flux-trap quantum coherent loops, similar to the case of non-magnetic mesoscopic conductors. The situation is different at lower fields, when fast spin dephasing occurs in the AFM and c-AFM regimes. At low field in the AFM regime, reproducible UCF traces and retraces are observed, after subtraction of a hysteresis due to the classical magneto-conductance background. Yet, as shown below from the correlation-field analysis, the nature of dephasing is different in the AFM regime. 
In the canted-AFM regime however, the UCF show a clear hysteresis, due to domain-wall pinning/depinning, with visible phase shifts and varying amplitudes upon the field-sweep direction, as also found for long conductors patterned from ultra-thin films \cite{Andersen2023}. This overall behavior of UCF is observed for all lengths, and their evolution is only correlated to the change in the conductor's magnetic state. 


To investigate dephasing and decoherence, we studied the UCF statistics for long conductors (Fig.~\ref{fig2}a), in the self-averaging limit, considering the temperature dependence both of the UCF correlation field $B_C$ (Fig.~\ref{fig2}b) and of their amplitude $\delta G^{rms}$ (Fig.~\ref{fig2}c). 
The correlation-field analysis is a powerful method to reveal the nature of dephasing. In a non-magnetic mesoscopic conductor, UCF have a quasi-period $B_C$ given by the ratio of the magnetic flux quantum $\Phi_0=h/(2e)$ to the largest size $S$ of coherent loops. For a magnetic field applied perpendicular to a 2D conductor $S=L_\varphi.L_\varphi$ ($L_\varphi<W$) or $S=L_\varphi.W$ ($L_\varphi>W$), where $W$ is the width of the nanostructure. 
Due to decoherence (that is, the decrease of $L_\varphi$ with temperature), $B_C$ increases with temperature when UCF are due to orbital dephasing only. In a mesoscopic magnet, this situation happens only for a uniform magnetization. For MnBi$_2$Te$_4$, this is achieved in large fields indeed, for $B_\perp>B_\perp^{sat}\approx7.5$~T (UM regime in Fig.~\ref{fig2}b). 
The situation is very different at low field, due to a spin dephasing mechanism induced by magnetism. The most prominent result is that $B_C$ has no temperature dependence, and it is significantly reduced with respect to its value in large fields, down to about 40mT in the AFM regime. Such temperature-independent fast UCF were also found in (Ga,Mn)As nanostructures, when dephasing is controlled by magnetism \cite{Granada2015}. This is not surprising since the magnetic properties do not vary much at low temperatures, well below the Néel temperature, so that the smaller $B_C$ value is not related to $L_\varphi(T)$ anymore but to the fast spin precession in exchange fields, e.g. due to pinned domain walls at the interface between AFM domains with opposite Néel vector \cite{Hedrich2021,Krizek2022}, or due to the spin splitting of local magnetic moments \cite{Jaroszynski1995}. 
In the c-AFM regime, for which the finite magnetization of canted domains can also modify the phase of coherent quasi-particles, a larger $B_C$ value is found, still with a negligible temperature dependence above 1K, when magnetic dephasing limits the UCF correlation field. Below 1K, large-enough flux-trap coherent loops start to contribute to faster UCF, so that $B_C$ decreases at lower temperatures in the c-AFM regime. 
Remarkably, independent of the magnetic state, the damping of the UCF amplitude with temperature is small (Fig.~\ref{fig2}c), a behavior at odds with that found in other mesoscopic magnets like (Ga,Mn)As \cite{Wagner2006,Vila2007}, and this quantum correction to the conductance remains important even at 4K. 
This is due to the specific transport properties of topological states in topological insulators, which have a reduced interaction with phonons and an enhanced transport mobility due to the anisotropic scattering by disorder \cite{Dufouleur2016}, 
and 
to the weak coupling of spin-textured Dirac states with their environment. 

To better understand the nature of quantum coherent transport, we further analyze the scaling laws of $\delta G_{rms}$ with the sample dimensions (width $W$ and length $L$). 
Analyzing the length dependence first (Fig.~\ref{fig2}d), the UCF amplitude follows a power-law dependence $\delta G_{rms}\propto 1/L^\alpha$ in the self-averaging limit, with a different value $\alpha^{A,B}$ for each sample (Fig.~\ref{fig2}d, inset). The average value $\alpha\approx$3 is twice larger than the value found in other mesoscopic magnets for quasi-1D coherent transport \cite{Vila2007}, which may result from the coexistence of bulk and topological surface states, with different $L_\varphi$ values. 
Considering the standard scaling analysis $\delta G_{rms}(T).L^\alpha \propto T^{-\beta}$, the slow decoherence is rooted into the small $\beta$ value ($\beta\approx0.25$), common to both devices, a much smaller value than found for (Ga,Mn)As nanostructures ($\beta\approx0.75$). Another difference with respect to previous studies of mesoscopic magnets is the influence of the aspect ratio of the conductor on the scaling analysis, which does not give a unique curve  
if only the length is considered. The geometric effect related to the width of the nanostructures is shown in Fig.~\ref{fig:S5}, with a better agreement to a universal law for the $W/L^\alpha$ scaling. 
Such an influence of the exact sample shape of the nanostructure on the scaling analysis is known for 2D mesoscopic conductors \cite{Lee1987}, with different scaling laws depending on the exact nature of dephasing, as due to inelastic scattering ($L_\varphi$) or to the thermal broadening of energy states (thermal length $L_T=\sqrt{\hbar D/k_BT}$). 
From the scaling analysis, we extract a value $L_\varphi \approx 500$~nm at $T=1$~K, in good agreement with the crossover between the self-averaging regime and the fully-coherent regime (vertical dashed line in Fig.~\ref{fig2}d, inset). This value of $L_\varphi$ is more than one order of magnitude longer that the state-of-the-art value found previously in (Ga,Mn)As mesoscopic magnets, for which decoherence is stronger (larger $\beta$ value). 
To confirm the nature of dephasing in MnBi$_2$Te$_4$, the thermal length $L_T$ must be compared to the phase-coherence length $L_\varphi$. 
For bulk samples, the typical value of 
the (Hall) transport mobility $\mu_b\approx$100cm$^2\cdot$V$^{-1}$s$^{-1}$ gives a transport length $L_{tr}\approx5$~nm and a thermal length $L_T\approx$100~nm at $T=1$~K, using values of the Fermi velocity $v_F\approx$5.5~ms$^{-1}$ and effective mass $m^*\approx0.15m_0$ determined by angle-resolved photoemission spectroscopy \cite{Otrokov2019,Chen2019,Hu2020}. 
For exfoliated flakes, the average value $\mu_f$ of the transport mobility in a nanostructure is larger than $\mu_b$, since bulk carriers coexist with topological surface states, due to the enhanced transport length of TSS resulting from anisotropic scattering \cite{Dufouleur2016}. The typical mobility ratio $\mu_{f}/\mu_b\approx5$ gives an average value of the transport length $L_{tr}\approx25$~nm and a corresponding thermal length $L_T\approx$240~nm at $T=1$~K. A more accurate assessment of the enhanced transport length of TSS is given by the measurement of their quantum mobility by Landau-level spectroscopy, which gives an elastic mean-free path $l_e\approx11$~nm \cite{Wissmann2025}. Considering the anisotropic scattering of 2D spin-textured Dirac fermions \cite{Culcer2010,Dufouleur2016}, with a ratio $L_{tr}/l_e$ close to a factor ten, a more reasonable value of the transport length for TSS is $L_{tr}\approx100$~nm, giving a corresponding thermal length $L_T\geq$500~nm at $T=1$~K. These values of the backscattering (transport) lengths are compatible with the 2D limit for charge transport, although transverse quantum confinement may play a role for device A, and the rather large diffusion constant $D^{TSS}\approx3\cdot10^{-2}$m$^2$/s confirms that thermal dephasing is probably negligible for TSS at low temperatures. 





\textbf{Contribution of topological surface states}

Strikingly, a direct evidence of the dominant contribution of 2D topological surface states to quantum interference is given by considering the angular dependence of UCF (Fig.~\ref{fig3}), which reveals the strong suppression of the UCF amplitude when the uniform magnetization is aligned within the sample plane. 
To do so, we compare the UCF measured for two different orientations of the magnetic field, applied either perpendicular to (OOP) or within (IP) the sample plane (Fig.~\ref{fig3}c,d, respectively). The evolution of the micromagnetic state is different for OOP and IP configurations. As seen in Fig.~\ref{fig3}a) from the classical magneto-conductance background, the in-plane magnetic field induces directly a continuous rotation of magnetic domains into a canted-AFM state, without a spin-flop transition since the field is applied perpendicular to the anisotropy axis, which gives little driving force on existing domain walls after a first magnetization curve. Similar to the OOP configuration, the UCF correlation field in the IP configuration is different for the c-AFM and the uniform magnetization states, but $B_C$ values are increased. In the c-AFM regime, this may be due to a slower evolution of the magnetization, with pinned domain walls, or to an additional Berry-phase contribution to spin dephasing due to the rotation of magnetic domains \cite{Granada2015}. 
In the uniform magnetization regime, $B_C$ is only due to structural disorder, and the change of $B_C$ for different orientations of the applied field usually results from a geometric effect since the in-plane field $B_{\vert\vert}$ probes coherent loops with a maximum area given by $t.W$, instead of $L_\varphi.W$ in the OOP configuration. This is a simple method to infer the value of $L_\varphi$ from the ratio $B_C^{OOP}/B_C^{IP}=L_\varphi/t$.
However, the situation is different for a magnetic topological insulator, due to the coexistence of 3D bulk carriers and of 2D topological surface states, the latter having an enhanced phase-coherence length in the diffusive regime \cite{Dufouleur2016}. 

As a consequence, the most remarkable result is the strong reduction of the UCF amplitude in the IP configuration, that is, when 2D-state coherent loops do not contribute anymore to field-induced conductance fluctuations. This is visible in Fig.~\ref{fig3}c,d, with a clear reduction of the UCF amplitude in the IP configuration, due to the shorter phase-coherence length of bulk carriers. 
As expected, the suppression of the TSS contribution to UCF is best observed for short-length conductors when giant-amplitude UCF arises from non-local contributions induced by TSS-related long coherent loops. This is best seen for device B (Fig.~\ref{fig3}b), the peak-to-peak UCF amplitude being six times smaller in the IP configuration, as compared to the OOP configuration. A similar behavior is found for device A, yet with a smaller relative change in the UCF amplitude, possibly due to the transverse quantum confinement of TSS in narrow nanowires, similarly to the case of non-magnetic 3D topological insulator quantum wires \cite{Dufouleur2013,Dufouleur2017}.




\textbf{Non-local conductance fluctuations at the micron scale}

The giant amplitude of UCF found for short-length conductors is a direct consequence of non-local contributions to quantum interference, which can result in conductance fluctuations that largely exceeds e$^2$/h, as known for non-magnetic mesoscopic conductors \cite{Benoit1987,Skocpol1987}. To confirm the specific nature of giant-amplitude UCF in MnBi$_2$Te$_4$ nanostructures, we studied quantum coherent transport in a non-local geometry for a uniform magnetization state in the OOP configuration, as shown in Fig.~\ref{fig4}.  
Non-local transport results in clear reproducible resistance fluctuations, measured for short distances (Fig.~\ref{fig4}c) but also some microns away from the current path (Fig.~\ref{fig4}d). Such a large non-local response, which is highly unusual for mesoscopic magnets, is favored by the long phase coherence length of topological surface states in intrinsic magnetic topological insulators, such as MnBi$_2$Te$_4$, combining a topological band structure with a large magnetization and strong anisotropy. This result strongly suggests that the inelastic scattering length is enhanced in intrinsic magnetic topological insulators, as compared to diluted magnetic topological insulators \cite{Mogi2017, Deng2022}, probably due to a larger magnetic gap and a more homogeneous magnetization.

\textbf{Conclusions}

In intrinsic magnetic topological insulators, the phase coherence length of topological surface states can reach the micron scale at low temperatures, resulting in unique quantum-coherent transport properties for a mesoscopic magnet, with significant non-local contributions to quantum interference and spin-related dephasing due to magnetism. Our results show that magnetization control at the nanoscale can directly tune interference patterns. 
Remarkably, the robust quantum coherence lifts some technological barrier that hampers since long the successful realization quantum devices based on magnets. Our finding opens a new route to investigate the quantum interference of spins in micron-sized nanomagnets with simple geometries, and to build a new type of quantum spin interferometers based on dephasing by tailored magnetic textures in the diffusive regime, different from spin-orbit based ballistic spin interferometers \cite{Datta1990}. 
Looking ahead, such magnetic textures would offer a natural platform for reconfigurable quantum interferometry. In particular, magnetic domain walls could be used as movable phase shifters, in addition to being useful for the dynamic routing of chiral transport \cite{Yasuda2017,Rosen2017}.



More generally, in the context of building novel quantum functional devices controlled by magnetism, our results contribute to the emergence of a new paradigm for quantum transport in magnetic topological insulators, based on the phase of spins, beyond topological quantum devices based on dissipationless edge states in the quantum anomalous Hall regime \cite{Gilbert2025}. Contrary to the study of Chern insulators, phase-coherent spin transport is not limited to ultra-thin nanostructures of magnetic topological insulators, and signatures of quantum interference persist at liquid-helium temperatures. 




\newpage


\textbf{Methods}

\textbf{Crystal growth}

Single-crystals of MnBi$_2$Te$_4$ were grown via the protocol developed in Refs.~\cite{Otrokov2019,Zeugner2019}, by slowly cooling down a stoichiometric mixture of molten Bi$_2$Te$_3$ and solid MnTe. The crystal structure and chemical composition of the crystals were validated by X-ray single-crystal diffraction and energy-dispersive X-ray spectroscopy. The obtained crystals are not air-sensitive, and they can be mechanically-exfoliated in normal atmosphere conditions.

\textbf{Nanofabrication and measurement setup}

Mesoscopic Hall bars of monocristalline MnBi$_2$Te$_4$ were obtained by the lateral patterning of thin flakes exfoliated in air from high-quality single crystals, with a thickness $t$ smaller than about 100~nm. After exfoliation, an ultra-thin TiO$_x$ hard mask was prepared by electron-beam lithography, metal evaporation and oxidation, and then used to mill a nano-Hall bar by using an Argon plasma. Ohmic contacts were finally prepared by electron-beam lithography and metal evaporation. In order to evidence the influence of the sample planar geometry on self-averaging, the universal conductance fluctuations were investigated in two devices with a different cross-section (device A : width $W=100$~nm, thickness $t=110$~nm; device B : $W=500$~nm, $t=55$~nm), each studied with three segments of different lengths (ranging from 200~nm to 3.5~µm; see Fig.~\ref{fig1}a,b) for details). The samples were mount onto the cold finger on an Oxford MX400 $^3$He/$^4$He dilution refrigerator, inserted into the bore of a 15~T magnet, and magneto-transport measurements were performed by using low-noise lock-in amplifiers, with low-enough signals so as to avoid electronic heating.


\textbf{Acknowledgments}

This work was supported by the European Union’s H2020 FET Proactive project TOCHA
(No. 824140), as well as by the CNRS International Research Project``CITRON".

\newpage

\appendix



\bibliography{MBTmeso}

\newpage

\begin{figure}[!h]
\includegraphics[height=\columnwidth]{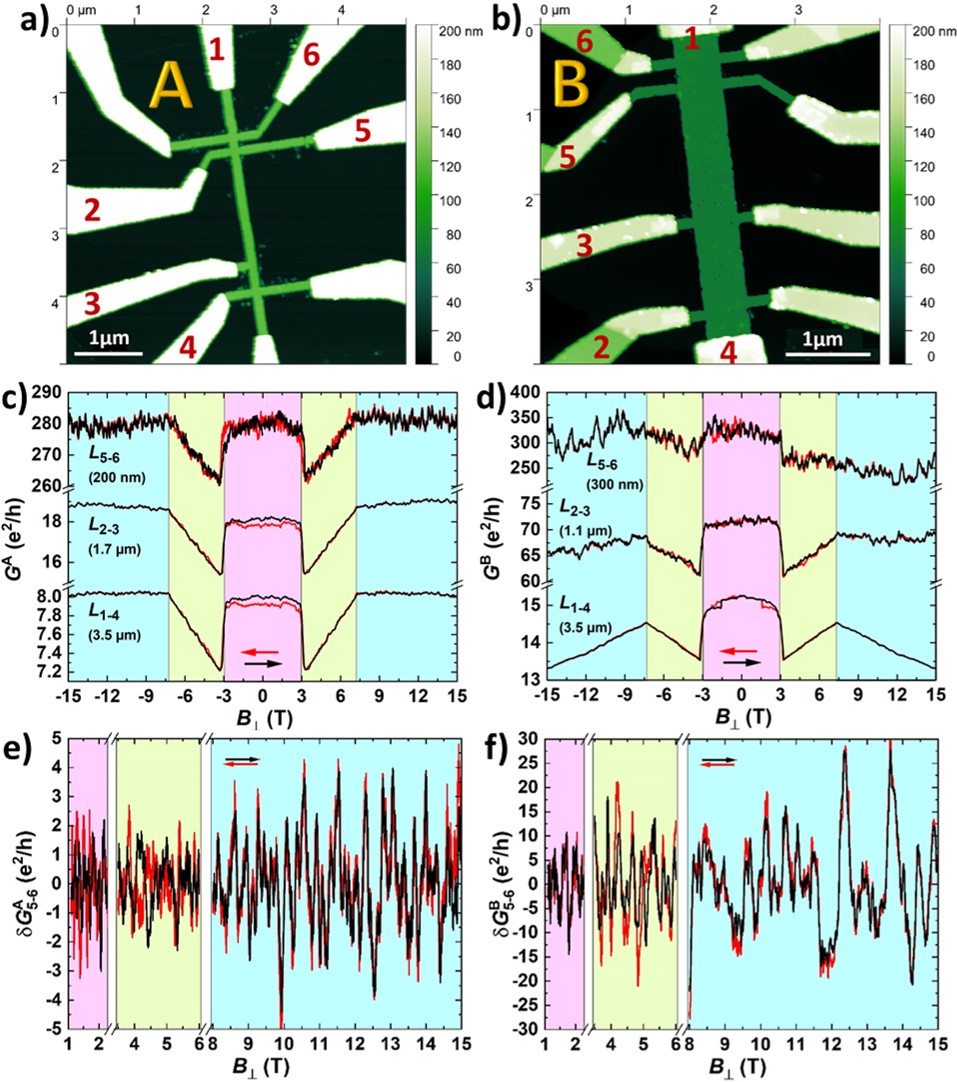}
\centering
\caption{\textbf{Giant conductance fluctuations in MnBi$_2$Te$_4$ nanostructures. }
\textbf{a,b}), AFM images of Hall-bar shaped devices A and B, with Ti/Au ohmic contacts. These two devices have a different cross-section (device A : width $W=100$~nm, thickness $t=110$~nm; device B : $W=500$~nm, $t=55$~nm), and conductance fluctuations are studied for three different lengths of mesoscopic conductors ($L_{14}=3500$~nm, $L_{23}=1700/1100$~nm, $L_{56}=200/300$~nm, for A/B respectively);  
\textbf{c,d}), Magneto-conductance measured at $T=1$~K, for three different lengths, with a magnetic field $B_\perp$ applied perpendicular to the sample plane (out-of-plane OOP configuration). The three different magnetic states are identified by colored panels (antiferromagnet - AFM, pink; canted antiferromagnet - c-AFM, green; saturated magnetization, aligned along the applied field, blue).;  
\textbf{e,f}), Giant-amplitude universal conductance fluctuations $\delta G_{5-6}$ for short-length conductors, after subtraction of the classical magneto-conductance.}
\label{fig1}
\end{figure}

\newpage

\begin{figure}[!h]
\includegraphics[width=\columnwidth]{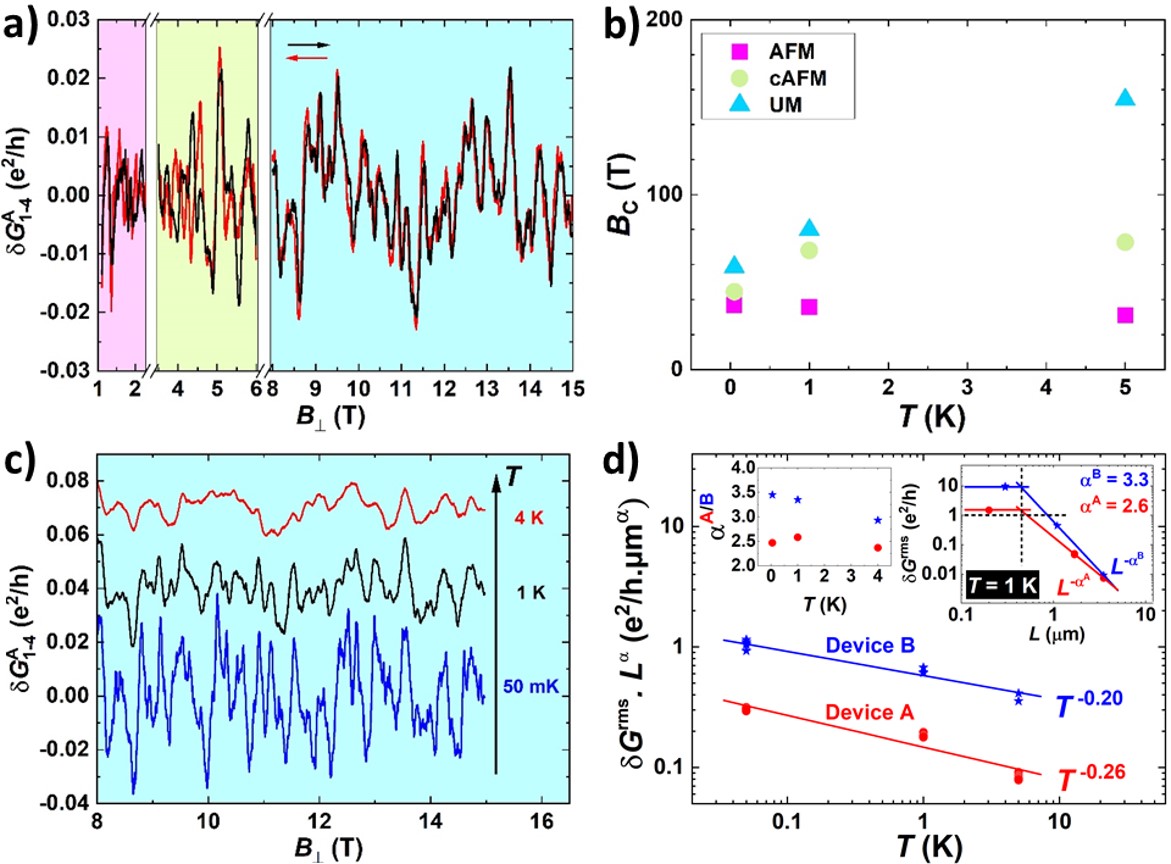}
\centering
\caption{
\textbf{Dephasing mechanisms and weak decoherence : correlation field, self averaging and scaling analysis.}
\textbf{a}), Magnetic field dependence of the universal conductance fluctuations $\delta G_{1-4}^A$, as measured at $T=1$~K for the 3.5~µm-long conductor in device A, after subtraction of the classical magneto-conductance;
\textbf{b}), Temperature dependence of the UCF amplitude $\delta G_{1-4}^A$ due to self-averaging ($L_\varphi < L_{1-4}$), showing the weak damping at higher temperatures;
\textbf{c}), Temperature dependence of the UCF correlation field $B_C$, measured for the different magnetic states. Whereas $B_C$ is directly related to the phase-coherence length in the uniform magnetization regime, the constant value in the AFM regime is due to spin dephasing; 
\textbf{d}), Scaling-law temperature dependence of the UCF amplitude $\delta G_{rms}.L^{\alpha}$ in the UM regime, revealing the slow decoherence ($\beta\approx0.25$). Inset (right): Length dependence of $\delta G_{rms}$. In the long-length limit ($L\gg L_\varphi$), the reduction of the UCF amplitude due to self averaging follows an usual power-law dependence $\delta G_{rms}\propto L^{-\alpha}$, with $\alpha \approx 3$. Inset (left): Temperature-dependence of $\alpha$}
\label{fig2}
\end{figure}

\newpage

\begin{figure}[!h]
\includegraphics[width=\columnwidth]{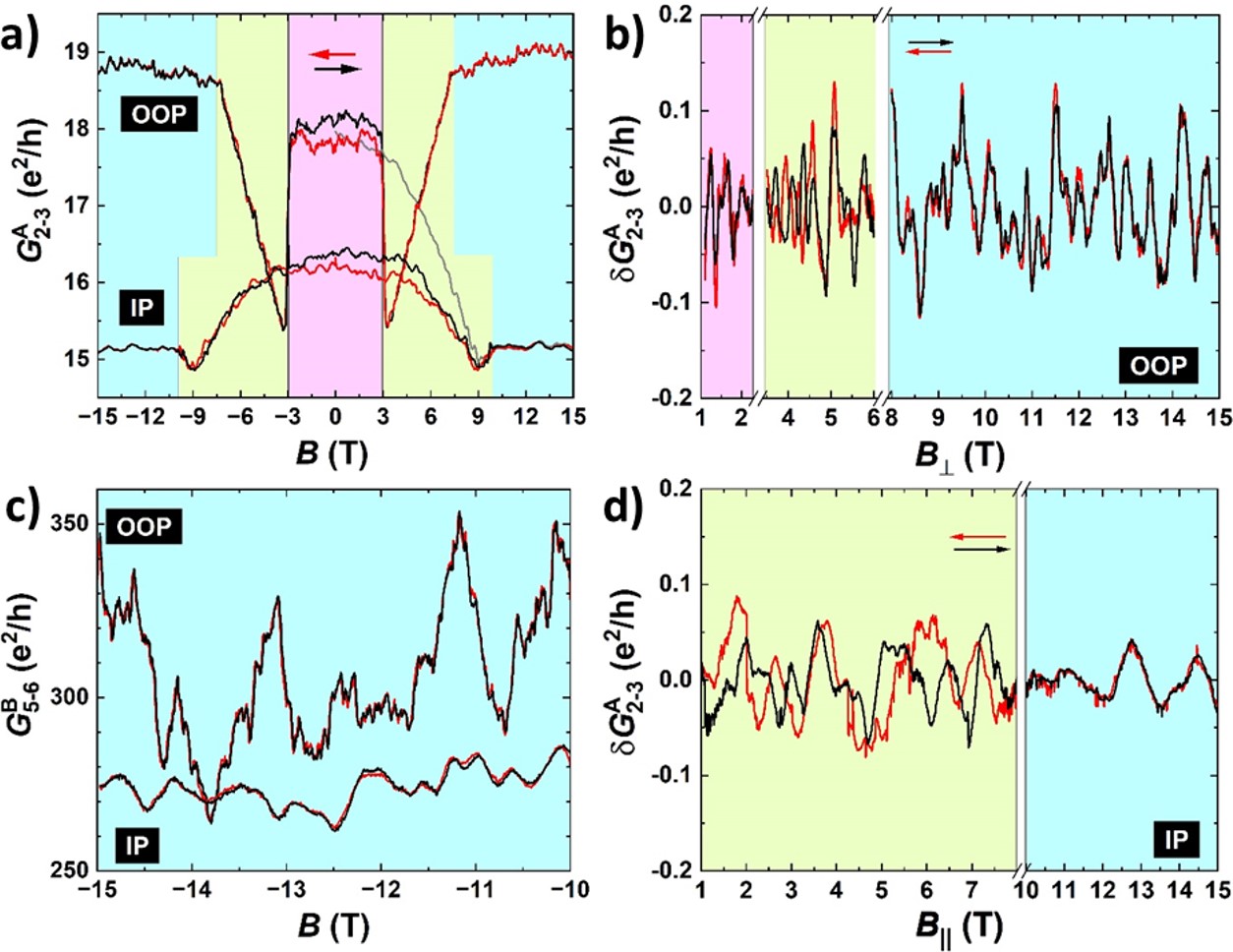}
\centering
\caption{ \textbf{Angular dependence and evidence for the TSS contribution to UCF.}
\textbf{a}), Magneto-conductance $G^{A}_{2-3}(B)$ measured at $T=1$~K in the OOP and IP configurations, with saturation fields $B^{sat}_{OOP}\approx7.5$~T and $B^{sat}_{IP}\approx10$~T. The grey trace is the first-magnetization curve measured in the IP configuration after cooldown in zero field;
\textbf{b,d}), Universal conductance fluctuations $\delta G^{A}_{2-3}(B)$, as measured in the OOP (b) and IP (d) configurations. The direct comparison in the uniform magnetization regime ($B>B^{sat}$) reveals the change in $B_C$ (expected from the geometry) and, 
strikingly, a significant change in the UCF amplitude (not observed in other mesoscopic magnets).
\textbf{c}) Comparison of $\delta G^{B}_{5-6}(B)$ at high magnetic fields, for both the OOP and IP configurations, showing the strong reduction of the UCF amplitude for the in-plane configuration (surface loops not probed by $B_{\vert\vert}$). }
\label{fig3}
\end{figure}

\newpage

\begin{figure}[!h]
\includegraphics[width=\columnwidth]{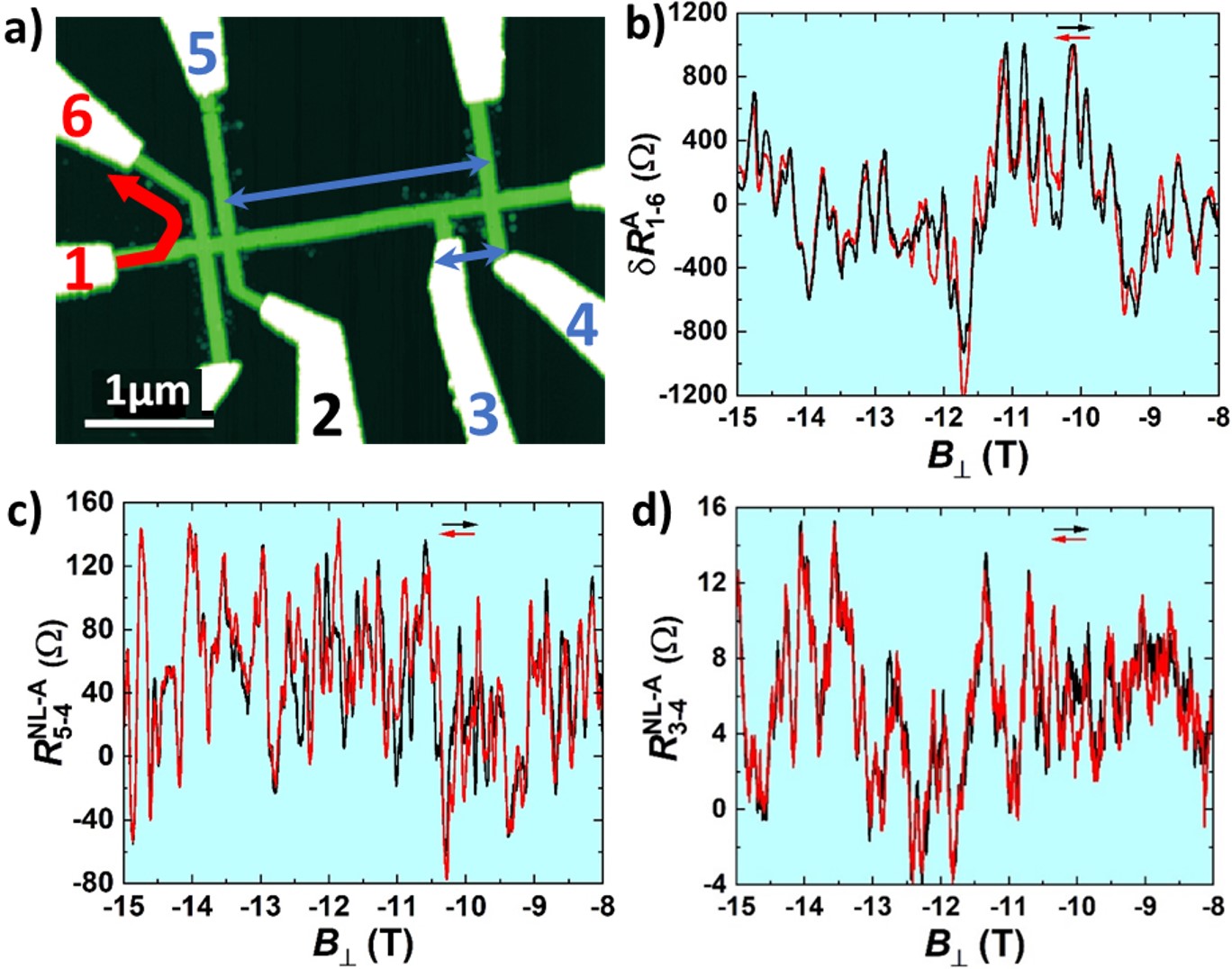}
\centering
\caption{ \textbf{Non-local measurements of quantum coherent transport at $T=$~1K.}
\textbf{a}), Current-injection path and voltage probes used for the non-local measurements of the magneto-conductance in Device A;  
\textbf{b}), Universal resistance fluctuations $\delta R_{1-6}^A$, as measured in the local configuration with a perpendicular field $B_{\perp}$, after subtraction of the classical magneto-resistance background;
\textbf{c}), Non-local magneto-resistance $R_{5-4}^{NL-A} (B_{\perp})$, representative of the UCF mostly probed by contact 5, about 200~nm away from the current path;
\textbf{d}), Non-local magneto-resistance $R_{3-4}^{NL-A} (B_{\perp})$, representative of the UCF probed by contacts 3 and 4, about 2µm away from the current path.}
\label{fig4}
\end{figure}

\clearpage
\setcounter{figure}{0}
\renewcommand{\thefigure}{S\arabic{figure}}

\onecolumngrid

\section*{Supplementary Information}
\subsection{Universal conductance fluctuations measured in the out-of-plane configuration at $T=$~1K, for devices A and B}
\begin{figure}[htbp]
    \centering
    \includegraphics[width=1\linewidth]{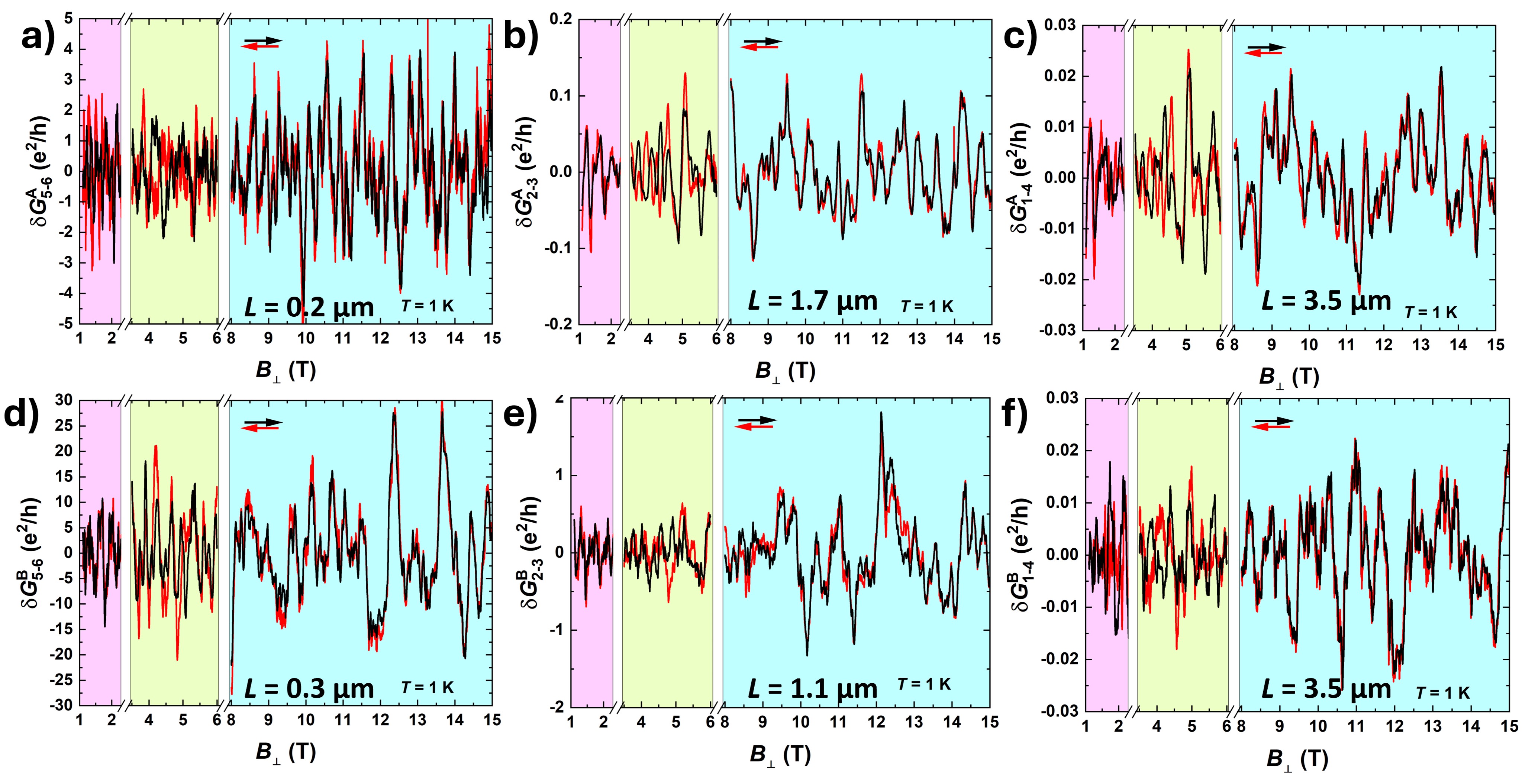}
    \caption{Universal conductance fluctuations measured under an out-of-plane magnetic field for a short-length $\delta G_{5-6}$, a medium-length $\delta G_{2-3}$ and a long-length $\delta G_{1-4}$ conductor at 1 K. Data shown for device A (\textbf{a,b,c}) and device B (\textbf{d,e,f}). Color code: antiferromagnetic (AFM) phase in pink, canted antiferromagnetic (cAFM) phase in yellow and phase of uniform magnetization (UM) in cyan. The perfect reproducibility of the UCF under both sweep directions in clearly seen in the UM phase, while in the cAFM phase the dephasing by local magnetic textures leads to some hysteresis.}
    \label{fig:S1}
\end{figure}
\newpage
\subsection{Universal conductance fluctuations measured in the in-plane configuration at $T=$~1K, for devices A and B}
\begin{figure}[htbp]
    \centering
    \includegraphics[width=1\linewidth]{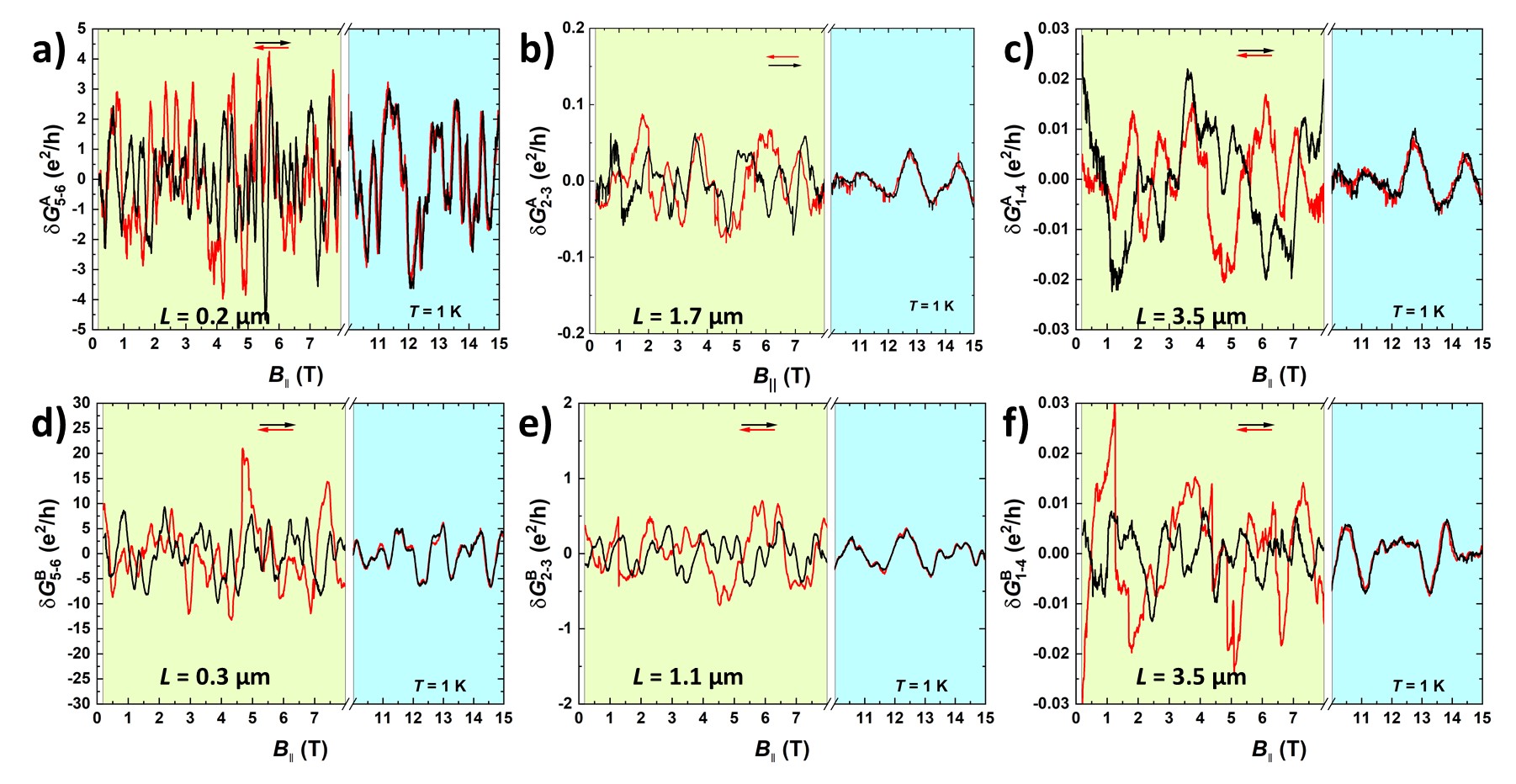}
    \caption{Universal conductance fluctuations under an in-plane magnetic field measured for all three lengths for device A (\textbf{a,b,c}) and device B (\textbf{d,e,f}). The correlation field $B_\mathrm{c}$ is significantly increased in the UM phase, with a visible larger quasi-period. Remarkably, the UCF amplitude is clearly reduced in the UM phase for the IP configuration, as compared to the UCF amplitude in the OOP configuration shown in Fig. \ref{fig:S1}.}
    \label{fig:S2}
\end{figure}
\newpage
\subsection{UCF correlation-field analysis}
\begin{figure}[htbp]
    \centering
    \includegraphics[width=1\linewidth]{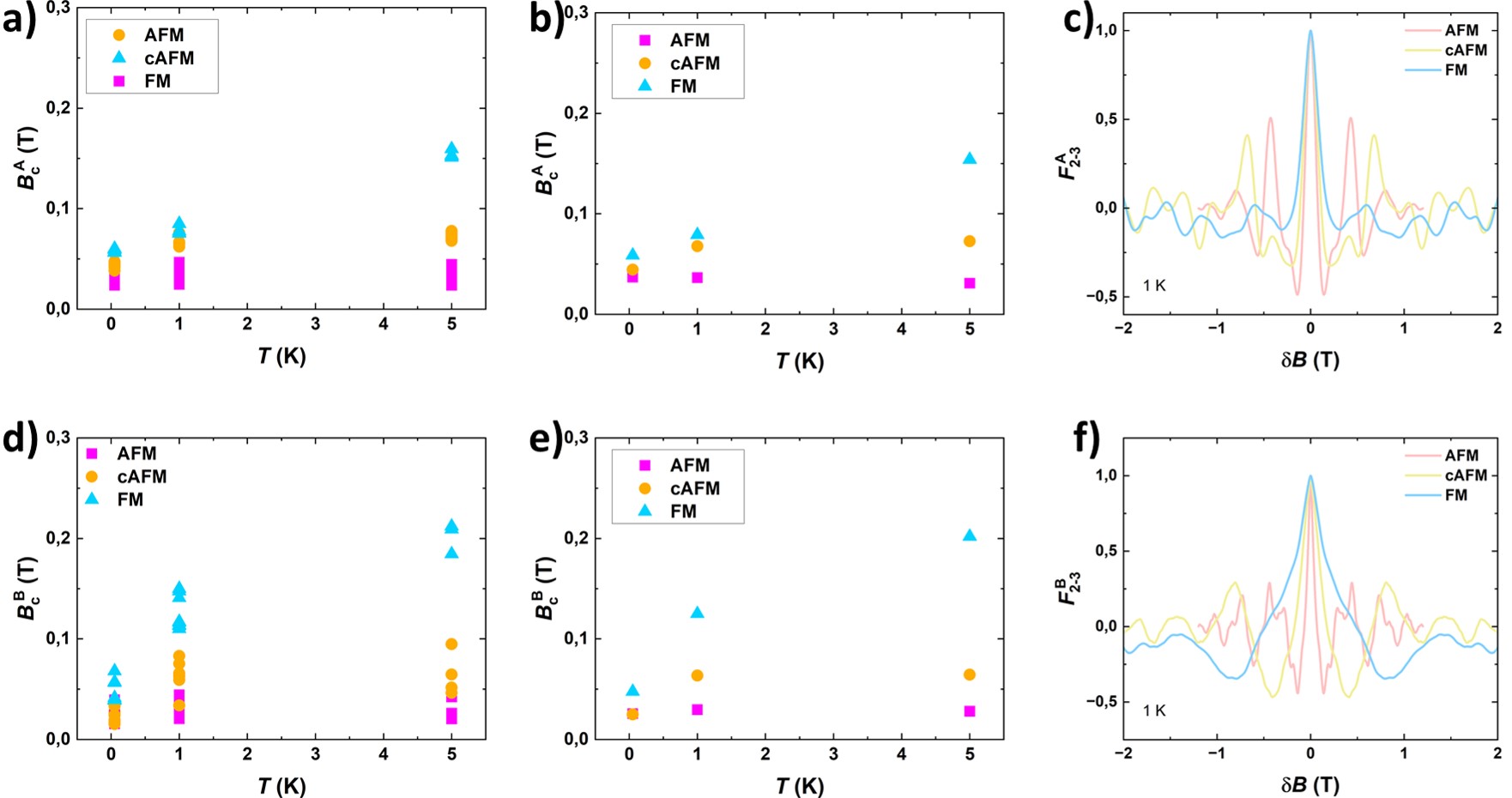}
    \caption{\textbf{a,d)} Values of the correlation field $B_\mathrm{c}$ extracted from the autocorrelation analysis for each conductor length, for device A and B, respectively. \textbf{b,e)} Average $B_\mathrm{c}$ values for all three magnetic phases, derived from data in a,d). \textbf{c,f)} Autocorrelation function $F(\delta B)$, normalized to $F(0)$, calculated for the medium-length conductor of device A (\textbf{c}) and device B (\textbf{f}), for the three different magnetic phases.}
    \label{fig:S3}
\end{figure}
The correlation field $B_\mathrm{c}$ is the field at which $G(B)$ and $G(B+B_\mathrm{c})$ are not correlated anymore, which corresponds to the UCF quasi-period $G(B)$ traces. $B_\mathrm{c}$ can be calculated using the autocorrelation function
\begin{align}
    F(\delta B) = \left< \delta G (B) \cdot \delta G (B+\delta B)\right>.\label{eq_autocorr}
\end{align}
The correlation field is defined by the half width at half maximum, as:
\begin{align}
    F(B_\mathrm{c})=\frac{F(0)}{2},
\end{align}
with $\sqrt{F(0)}=\delta G^\mathrm{rms}$.\\
Two exemplary autocorrelation plots at 1 K are shown in Fig. \ref{fig:S3} c,f). It is clearly visible that the width of $F(\delta B)$ is larger for the high-field phase of uniform magnetization (cyan) compared to the cAFM and AFM phases (yellow and pink, respectively). \\
The values of $B_\mathrm{c}$ shown in Fig.~\ref{fig2}b) and in Fig.~\ref{fig:S3} b,e) are the average values calculated from the 1-4 and 2-3 conductors for positive and negative fields and up- and down sweeps of fields, for each device.
\newpage

\subsection{Electronic heating}
\begin{figure}[htbp]
    \centering
    \includegraphics[width=1\linewidth]{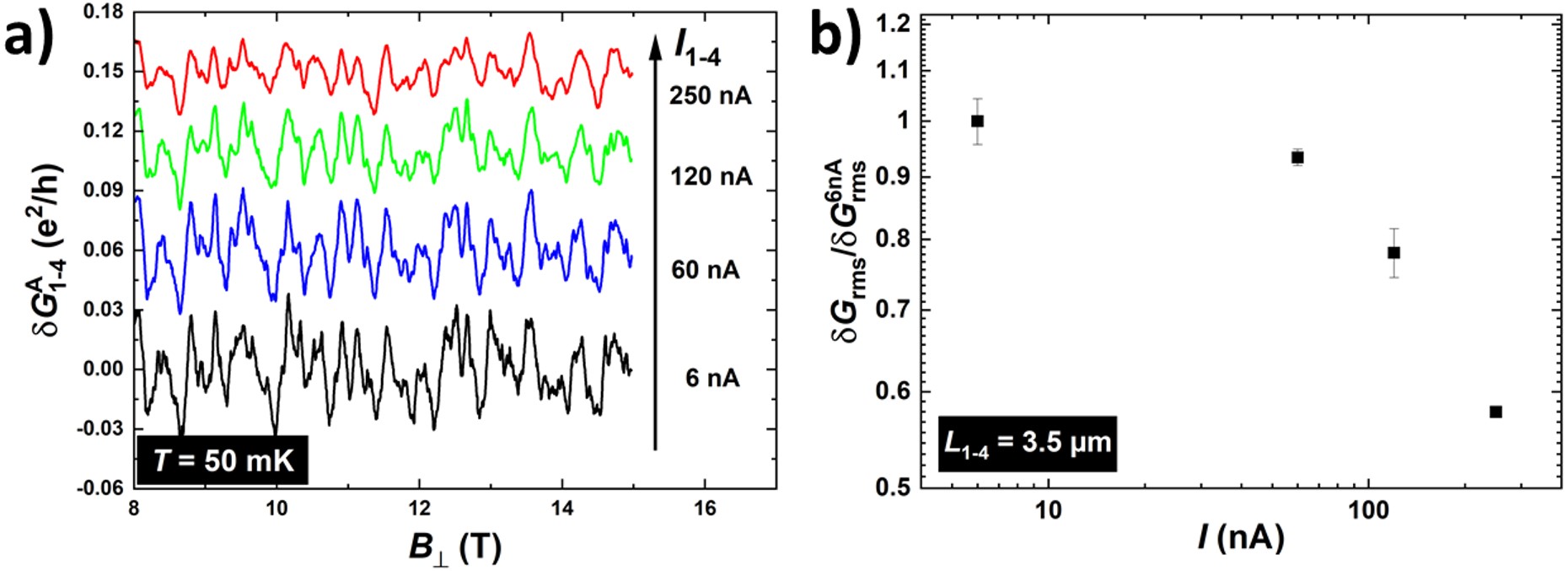}
    \caption{\textbf{a)} High-field dependence of the UCF measured with the longest conductor of device A, at base temperature for different currents $I$. Curves offset for clarity. \textbf{b)} Current dependence of the UCF amplitude $\delta G_{rms}$ measured at base temperature. 
    }
    \label{fig:S4}
\end{figure}

Quantum transport measurements were performed with the samples mount onto the cold finger of a  $^3$He/$^4$He dilution refrigerator, with a fixed position, either perpendicular or parallel to the magnetic field direction of a 15~T superconducting-coil magnet. To change from the OOP to the IP configuration, the samples were thermal cycled up to room temperature and cooled down in a second run. Data measured at $T=1$~K were obtained by operating the 1~K pot only. The dilution refrigerator has a lowest phonon-bath temperature $T_{base}=30$~mK, but a little higher electronic temperature at base, around 50~mK. For mesoscopic transport measurements, low-enough current levels were used to avoid electronic heating, which otherwise happens at very low temperatures if $eV>k_BT$, leading to a significant increase of the electronic temperature for large currents ($T_{elec} > T$). This effect is best evidenced by considering the long conductor (large resistance), which results in the decrease of the UCF amplitude for too-high current levels (Fig.~\ref{fig:S4}a,b). For small-enough currents, the UCF amplitude is independent of the current (plateau behavior). In general, measurements are performed with the maximum current level that corresponds to the departure from the plateau (about 60~nA in Fig.~\ref{fig:S4}b, for $T=50$~mK), so as to optimize the signal-to-noise ratio. 


\newpage\subsection{Scaling-law analysis : influence of the aspect ratio $W/L$}
\begin{figure}[htbp]
    \centering
    \includegraphics[width=1\linewidth]{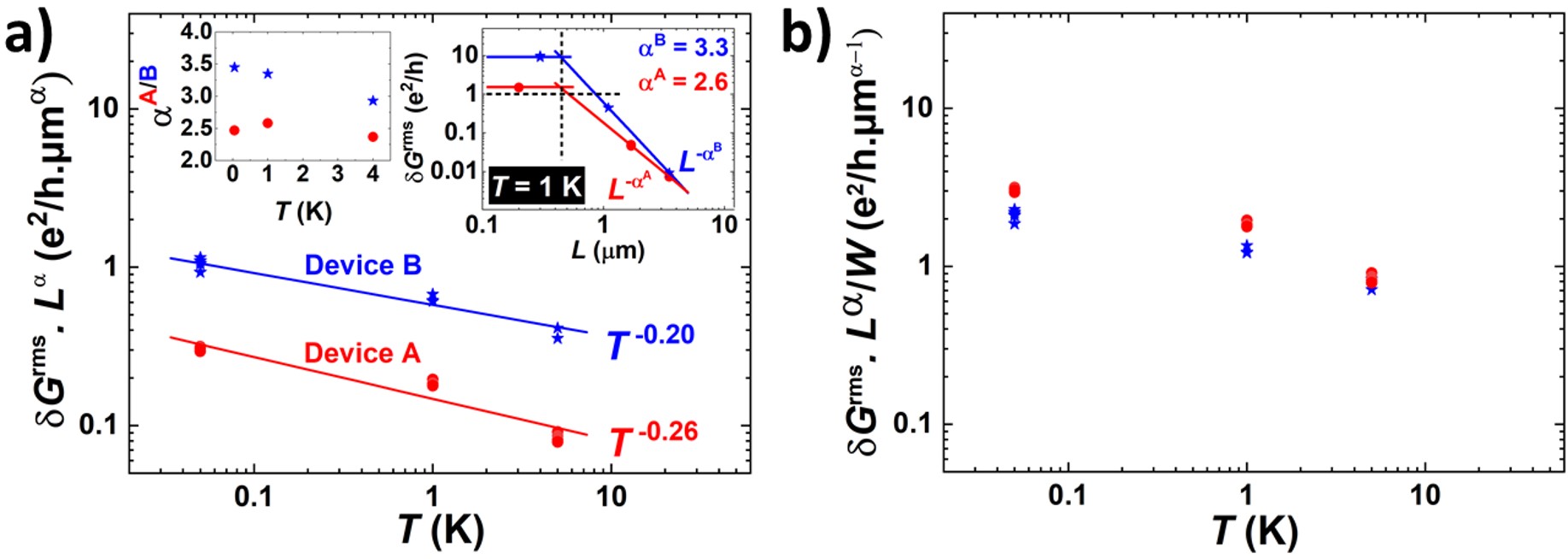}
    \caption{Scaling of $\delta G_\mathrm{rms}(T)$ considering (a) the length dependence only (same figure as Fig.~\ref{fig2}d), or (b) both the length $L$ and the width $W$ dependence. The UCF amplitude is measured for long conductors ($\delta G_{2-3}$ and $\delta G_{1-4}$, in the self-averaging regime) in large OOP magnetic fields (UM phase).}
    \label{fig:S5}
\end{figure}

\end{document}